\documentclass[a4paper]{spie}  %>>> use this instead for A4 paper
\pdfoutput=1
\usepackage{amsmath,amsfonts,amssymb}
\usepackage{graphicx}
\usepackage[colorlinks = true,linkcolor = blue,citecolor = blue]{hyperref}

\usepackage{aas_macros}
\usepackage[caption = false]{subfig}

\newcommand{\astrosat}{AstroSat}

\title{In-orbit performance of {\astrosat} CZTI}

\author[a]{Santosh V. Vadawale}
\author[b]{A. R. Rao}
\author[c]{Dipankar Bhattacharya}
\author[c]{Varun B. Bhalerao}
\author[c]{Gulab Chand Dewangan}
\author[c]{Ajay M. Vibute}
\author[a]{Mithun N. P. S.}
\author[a]{Tanmoy Chattopadhyay}
\author[d]{S. Sreekumar}
\affil[a]{Physical Research Laboratory, Navarangpura, Ahmedabad, India}
\affil[b]{Tata Institute of Fundamental Research, Homi Bhabha Road, Colaba, Mumbai, India}
\affil[c]{Inter-University Center for Astronomy and Astrophysics, Pune, India}
\affil[d]{Vikram Sarabhai Space Center, Thiruvanathapuram, India}

\authorinfo{Correspondence E-mail: santoshv@prl.res.in, Telephone: 079-2631 4616}

\begin{document}
\maketitle

\begin{abstract}
Cadmium-Zinc-Telluride Imager (CZTI) is one of the five payloads on-board 
recently launched Indian astronomy satellite {\astrosat}. CZTI is primarily 
designed for simultaneous hard X-ray imaging and spectroscopy of celestial 
X-ray sources. It employs the technique of coded mask imaging for measuring 
spectra in the energy range of 20 - 150 keV. It was the first scientific 
payload of {\astrosat} to be switched on after one week of the launch and 
was made operational during the subsequent week. Here we present preliminary 
results from the performance verification phase observations and discuss 
the in-orbit performance of CZTI. 
\end{abstract}


%%%%% INTRODUCTION %%%%%%%%%%%%%%%%%%%%%%%%%%%%%%%%%%%%%%%%

\section{Introduction}
\label{intro}

\begin{figure}
\begin{center}
\includegraphics[width=0.7\textwidth]{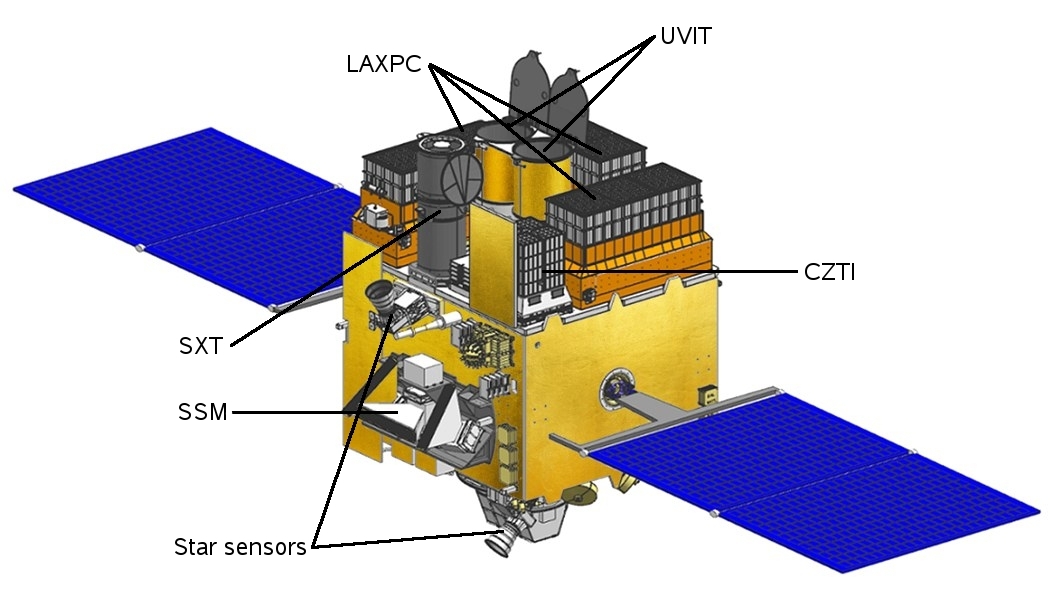}
\caption{
	\label{astrosat_deployed}
	Deployed configuration of {\astrosat} depicting the locations of 
various scientific instruments (image credit ISRO).}
\end{center}
\end{figure}

{\astrosat} is India's first dedicated satellite mission for multi-wavelength 
Astronomy~\cite{singh_astrosat}. It was launched on September 28, 2015 
by Polar Satellite Launch Vehicle PSLV C-30 of the Indian Space Research 
Organization (ISRO) into 650 km orbit with inclination of 6 degrees. 
{\astrosat} carries five instruments: (i) Large Area Xenon Proportional 
Counter (LAXPC), (ii) Soft X-ray Telescope (SXT), (iii) Cadmium Zinc 
Telluride Imager (CZTI), (iv) Ultra-violet Imaging Telescope (UVIT) 
and (v) Scanning Sky Monitor (SSM). The first four instruments are co-aligned 
to provide simultaneous multi-wavelength observations of astrophysical 
sources. The SSM is mounted on a rotating platform and scans the sky 
for detecting new transient X-ray sources as well as to monitor the 
variability of known X-ray sources. {\astrosat} also carries an auxiliary
instrument - Charge Particle Monitor (CPM) to detect the presence of 
South Atlantic Anomaly (SAA). 
Figure~\ref{astrosat_deployed} shows the location of all 
instruments of {\astrosat} in deployed condition. 

All instruments on-board {\astrosat} as well as 
all satellite subsystems are functioning normally since the launch. The
period of first six months after launch was designated as performance
verification (PV) phase, which was successfully completed in March 2016.
The observations during the PV phase were used to carefully characterize 
all instruments and to obtain actually realized values for various 
instrument parameters. The period of next six months (i.e. till September
2016) is reserved for the scientific observations by the instrument teams, 
which is successfully progressing at present. From October 2016 onwards,
{\astrosat} will be available for guest observations, initially for the 
Indian astronomers for one year, and subsequently open to all. During 
this period, {\astrosat} will be operated as a proposal driven observatory. 
The nominal lifetime of {\astrosat} is planned to be five years. Thus it 
will provide a great opportunity to carry out observations of a variety of
astrophysical sources in wavebands ranging from visible to hard X-rays.

Here we present the results of in-orbit characterization of the CZTI
instrument carried out with the observations during the PV phase. 
Section~\ref{inst_desc} provides the detailed description of the instrument, 
a brief description of pre-flight calibration is presented in section~\ref{calib}, 
section~\ref{inorbit} discusses the in-orbit operations and in section~\ref{results} preliminary results 
from CZT Imager are presented. 

%%%% Instrument details %%%%%%%%%%%%%%%%%%%%%%%%%%%%%%%%%%%%%%%%%%%

\section{CZTI instrument description}

\label{inst_desc}
\begin{figure}
   \begin{center}
   \begin{tabular}{c c c} %% tabular useful for creating an array of images 
   \subfloat[]{\label{czti_full} \includegraphics[width=0.25\textwidth]{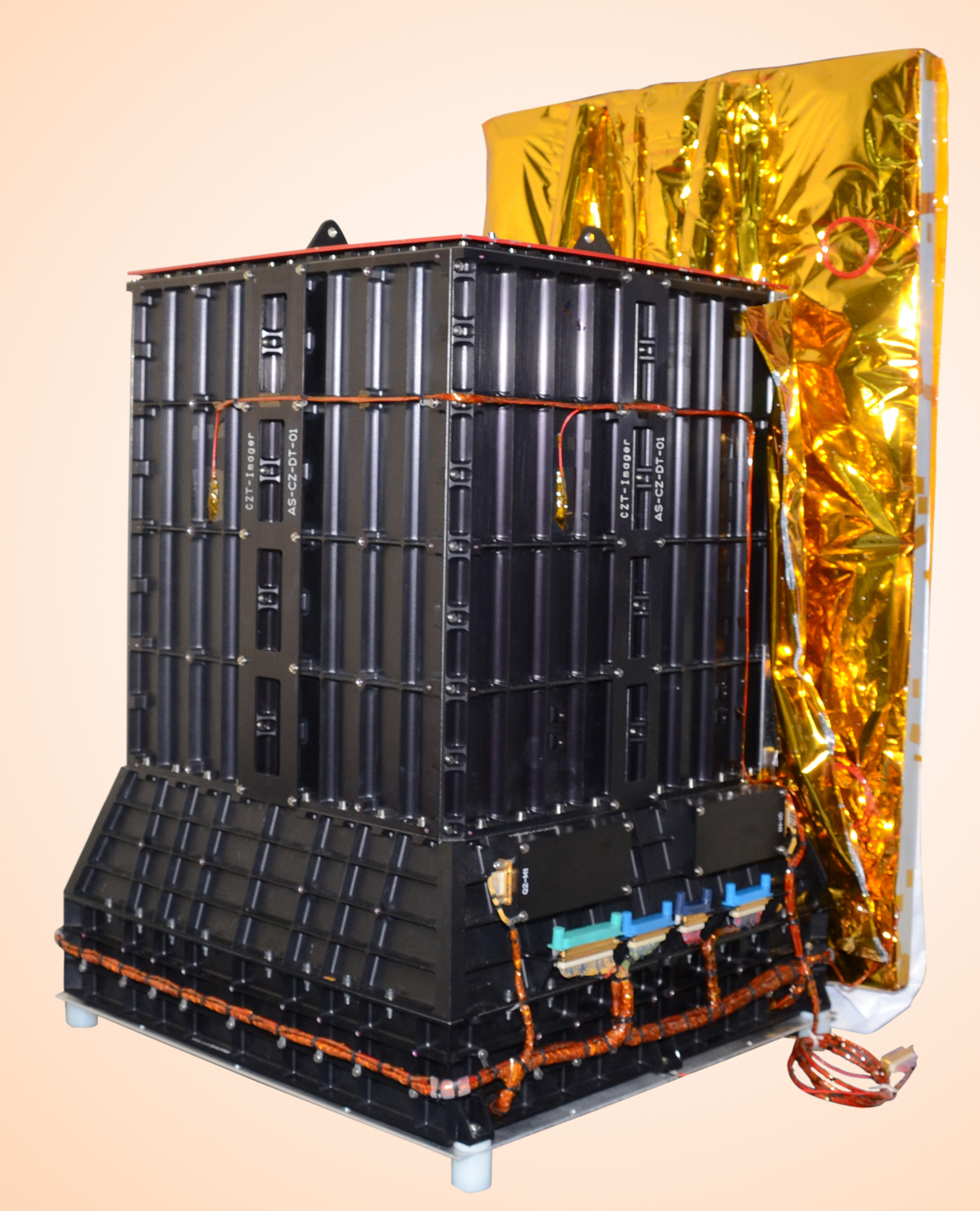}}
   \subfloat[] {\label{czti_quad} \includegraphics[width=0.30\textwidth]{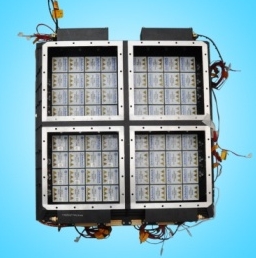}}
   \subfloat[]{\label{czti_cam} \includegraphics[width=0.35\textwidth]{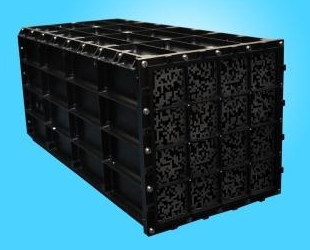}}
   \end{tabular}
   \end{center}
   \caption[example]
   { \label{czti_instr}
    CZT Imager: (a) Fully assembled CZTI payload. (b) CZTI detector plane 
with 64 detector modules arranged to four identical quadrants. (c) Collimator
with Coded aperture mask for one quadrant. 
}
\end{figure}

Cadmium Zinc Telluride Imager is a coded aperture telescope with the 
primary objective of simultaneous imaging and spectroscopy of bright
X-ray sources in the hard X-ray band (20-150 keV). It consists of 
a large area, pixelated detector plane and a two dimensional coded 
aperture mask (CAM) placed at a distance of 478 mm from the detector 
plane. It has a total active area of 976 $cm^2$ which is achieved by 
an array of 64 pixelated CZT detector modules. Each detector module
is of size $39.06~mm~\times~39.06~mm$ and has a $16~\times~16$ pixel array. 
Each pixel is of size $2.46~mm\times~2.46~mm$ (except the edge and corner pixels which are of slightly smaller dimensions), 
resulting in a total of 16384
pixels.The whole CZTI instrument
is designed as four independent but identical quadrants, each having 
$4~\times~4$ array of the CZT detector modules and associated coded
aperture mask. Figure~\ref{czti_full} shows the fully assembled CZTI 
whereas figures ~\ref{czti_quad} show the CZTI 
detector plane in the form of four identical quadrants. 
%
%Coded Aperture Mask (CAM) is employed for indirect imaging and 
%simultaneous background measurement. At higher energies ($>$100 keV)
%CZTI acts as an open detector as the collimators becomes progressively 
%transparent. 
%
%\textcolor{red}{
%Primary science objectives include study of accreting black hole and neutron star sources and of transient events like gamma ray bursts.}
%

Passive collimators placed in each quadrant, restrict the field of view of 
the detector array to 4.67$^\circ$ x 4.67$^\circ$ at energies 
below 100 keV and support the coded aperture mask as shown in 
figure~\ref{czti_cam}.
The mask is made of 0.5 mm  thick Tantalum plate with open and closed 
square/rectangular elements of the same size as detector pixels. CZTI 
employs the box-type mask design where the size of the full mask is same 
as the detector plane in each quadrant. In such a design, detectors are 
exposed to the shadow of entire mask only for on-axis sources. At other 
off-axis locations, only a part of the shadow is recorded by the detectors. 
The mask patterns are based on 255 element pseudo-noise Hadamard set URAs 
(Uniformly Redundant Arrays). Seven of sixteen such possible patterns were 
chosen based on the considerations of mechanical support to each element.
Each pattern which has an array of $16~\times~16$ mask elements was used as coded mask for 
individual detector module and a random arrangement of these into $4~\times~4$ 
array yielded the pattern for one quadrant. Mask patterns for other three 
quadrants were obtained by rotation of this mask pattern by 90$^\circ$, 
180$^\circ$ and 270$^\circ$. One important aspect of the collimators 
and mask is that these are designed to be effective up to 100 keV and hence 
become progressively transparent beyond 100 keV. However, the CZT detectors 
themselves have significant detection efficiency beyond 100 keV. This 
enables CZTI to act as an all sky detector at higher energies resulting 
in the enhanced capability of CZTI as GRB detector as well as hard X-ray 
polarimeter, as discussed in subsequent sections. 

The CZT detector modules, procured from Orbotech medical solutions, 
operate at near room temperatures. Each CZT detector module has 
256 pixels which are read out by two Application Specific Integrated Circuits 
(ASIC) integrated within the module. Each ASIC has 128 channels consisting
of analog chain of preamplifier, shaper, discriminator followed by 
self-triggering and interface circuitry. For each incident X-ray photon 
with energy above a set lower threshold, pixel address and pulse height 
are stored within the ASIC and are subsequently read by the Front-end
Electronic Board (FEB) which assigns time stamps to the recorded events.
The detector modules are mounted on a thermally conductive plate 
which is connected to the external radiator plate through heat pipes, so 
as to maintain the detector modules at temperature in the range of 
5$^\circ$C - 15$^\circ$C. 
%Each quadrant has its own Front-end Electronic 
%Board (FEB) which acquires the data from the detector modules and assigns 
%time stamps to events. 

Each CZTI quadrant has a 20 mm thick scintillator detector (CsI(Tl)) 
placed below the CZT detector plane, to act as a veto detector. The 
scintillator is read by photo multiplier tubes from two sides. High energy 
particles will interact with CZT detectors and then with CsI detector 
producing simultaneous signals allowing to filter out such events. Each 
quadrant also has one calibration module placed between the detector plane
and base of the collimators. It contains $^{241}Am$ radioactive source 
embedded in a cesium iodide crystal. The $^{241}Am$ source is a standard 
source used for calibration of hard X-ray detectors with its mono-energetic 
X-rays of 59.54 keV. However, the emission of each photon of 59.54 keV 
in this source is also associated with the simultaneous emission of one alpha 
particle, which is absorbed by the CsI crystal being read by a photo diode.  
Thus the event in the CZT detector corresponding to the 59.54 keV photon
can be easily identified by simultaneous detection of the alpha particle~\cite{rao_alpha}. 
These alpha tagged photons are used for continuous 
in-flight calibration of CZT detectors.

All four quadrants are controlled by a common Processing Electronics (PE) 
of CZTI which also controls all interfaces with the spacecraft. Event data 
acquired from the 16 detector modules of each quadrant are packaged by 
the quadrant wise FEB and sent to the PE every second, which then further 
re-packages them into larger data frames and sends to spacecraft for 
on-board storage. Other functions of the FPGA based PE include distribution 
of power from the satellite to detector quadrants, forwarding telecommands to 
the detectors, identifying the mode of operation based on telecommand or 
environment and acquiring the house keeping parameters. 
PE also has an 
interface with a GPS based on-board satellite positioning system (SPS). 
The SPS provides a pulse every 16 seconds with accuracy of 200 ns, which 
is used for absolute time assignment for individual CZTI events.
%(IN MY UNDERSTANDING, THIS IS NOT CORRECT. STBG of laxpc is interfaced with 
%sps and czti pe and other instruments' PE is interfaced with STBG. 
%STBG provides the pulse.)
%}
%\textcolor{blue}{
%PE also has an interface to STBG of LAXPC, which provides pulse every 16 seconds
%which is used to latch CZT clock with Universal Time.}
Though {\astrosat} is in a low inclination orbit, it has passages through South 
Atlantic Anomaly (SAA) region of increased particle concentrations. The Charge 
Particle Monitor (CPM) on-board {\astrosat} measures particle flux and provides 
SAA signal to all X-ray instruments including CZTI when particle flux is 
above a certain threshold. CZTI processing electronics detects this SAA 
signal and issues commands to turn off detector HV. Control of overall
{\astrosat} spacecraft as well as data down link from all scientific instruments
is carried out from the ISRO Telemetry, Tracking and Command Network (ISTRAC) 
located at Bengaluru. The Payload data is then sent for processing at 
the payload operation centers (POC) for individual payloads and the
higher level data is archived by Indian Space Science Data Center (ISSDC) 
located at Bylalu. The payload operation center for CZTI is located at the 
Inter University Center for Astronomy \& Astrophysics (IUCAA) at Pune.

%%%%%%%%%%%%%%%%% Ground calibration %%%%%%%%%%%%%%%%%%%%%%%%%%%%%%%%%
\begin{figure}
\begin{center}
   \begin{tabular}{c c c} %% tabular useful for creating an array of images 
    \subfloat[]{\label{mutau_fit} \includegraphics[width=0.30\textwidth]{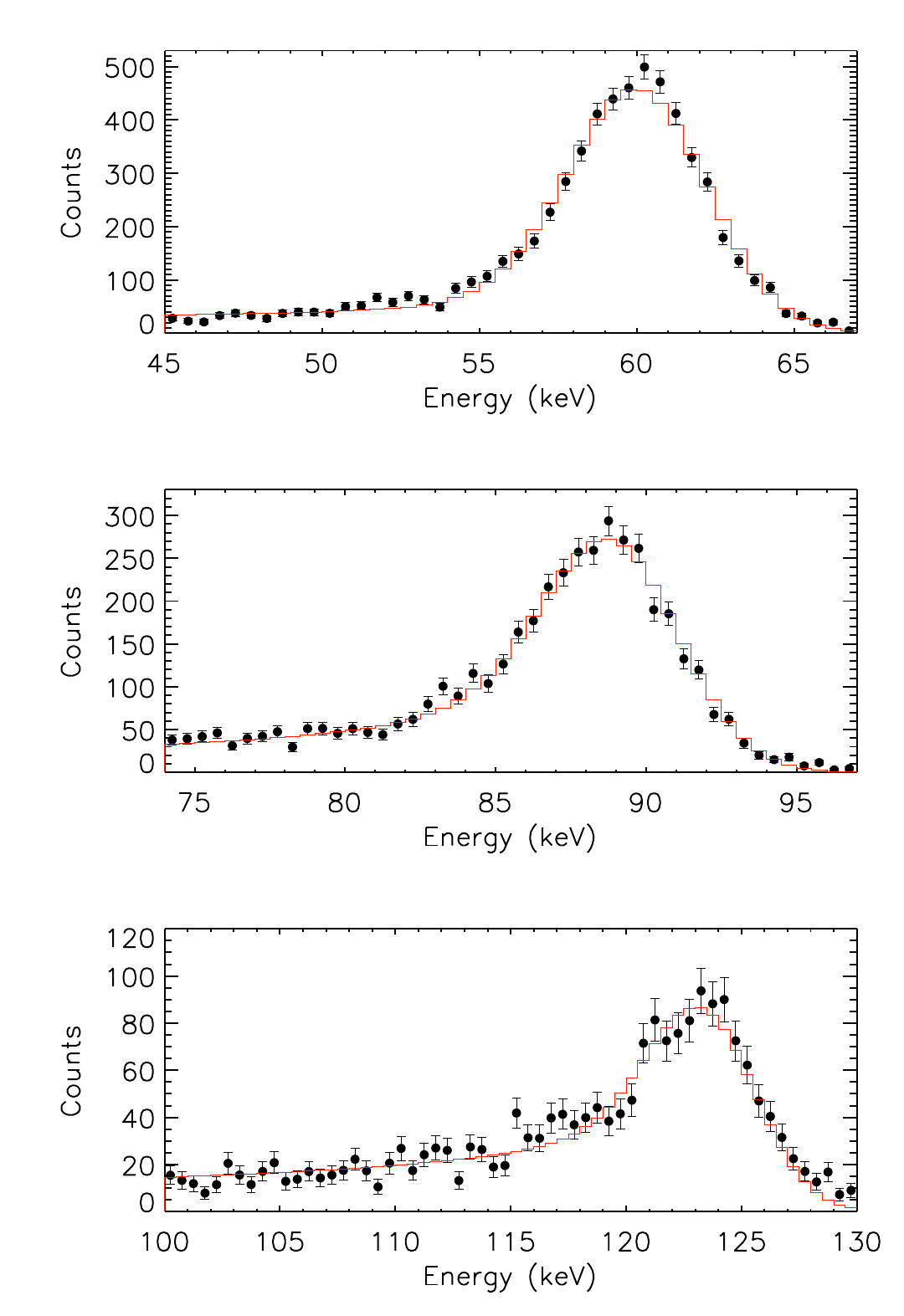}}
   \subfloat[]{\label{mutau_param} \includegraphics[width=0.62\textwidth]{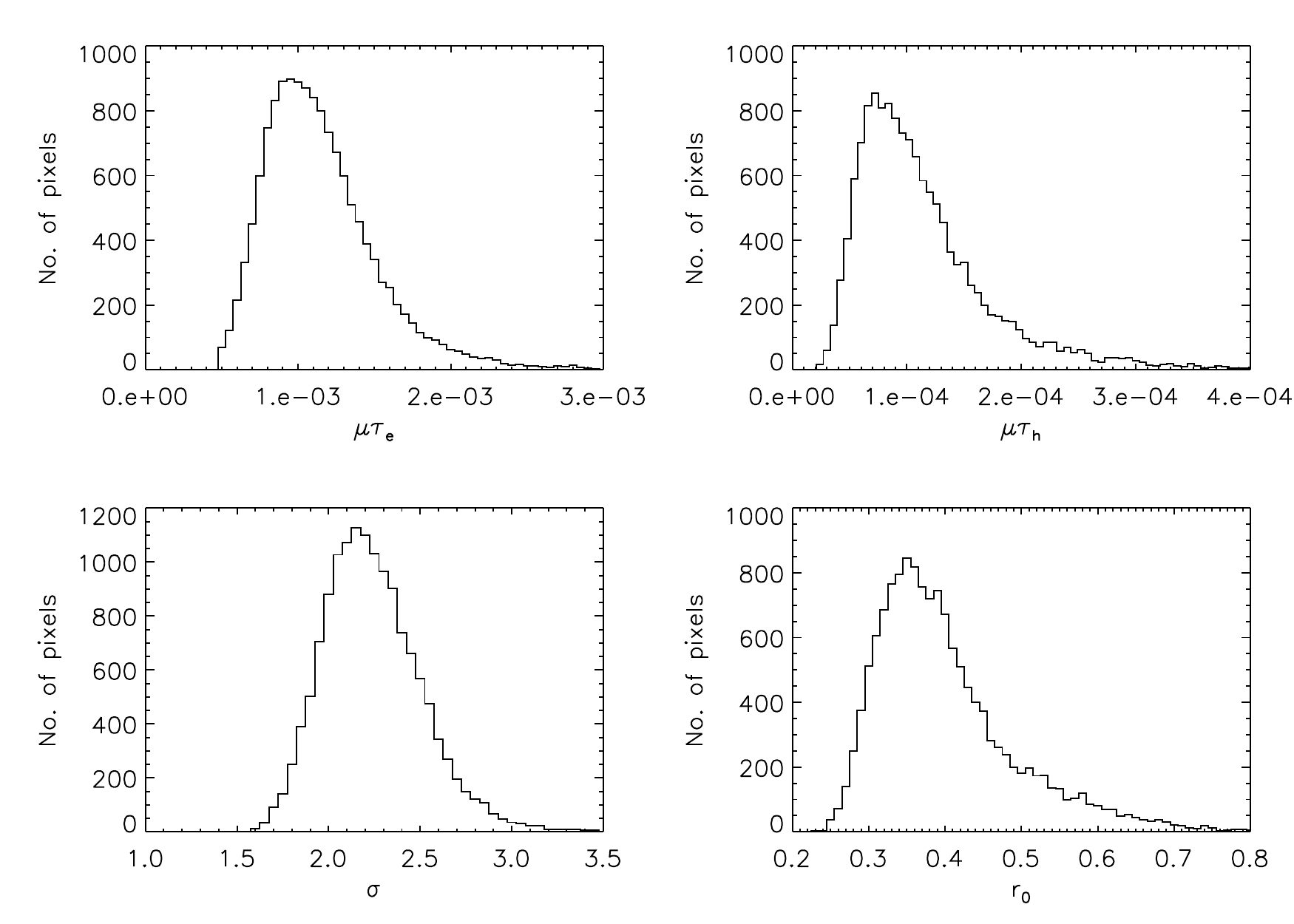}}
   \end{tabular}
\caption{(a) Simultaneous fit of the model to spectra at three energies (b)
Distribution of fit parameters for all pixels at 10$^{\circ}$ C temperature
}
\end{center}
\end{figure}

\section{CZTI pre-flight calibration}

\label{calib}
CZTI consists of total 16384 pixels and each pixel is essentially an 
independent X-ray detector. Thus each individual pixel must be properly
characterized in terms various parameters such as energy resolution,
relative efficiency, gain, offset etc. This massive exercise was carried
out at the Vikram Sarabhai Space Center of ISRO. 
%\textcolor{red}{\bf  where the full CZTI payload was fabricated (IS THIS CORRECT??, SHOULD WE REMOVE THIS)} 
Each quadrant was tested with three radioactive
sources 
$^{109}Cd$ (lines at 22.0 keV and 88.0 keV), $^{241}Am$ 
(line at 59.54 keV) and $^{57}Co$ (lines at 122.0 keV and 136.0 keV) at 
five temperates in the range of 0$^{\circ}$ C to 20$^{\circ}$ C with
a step of 5$^{\circ}$ C. 
Spectrum from each pixel was fitted for these energies in order to 
obtained pixel wise energy resolution, linearity, gain, offset; which 
were then stored in the calibration database (CALDB) of CZTI. 
It is well known that the spectrum of a mono-energetic line recorded
by CZT detector typically has a strong tail at low energies and thus 
can not be approximated by a Gaussian shape. Such an asymmetric line shape 
of the CZT detector can be explained as arising due to the charge 
trapping and can be quantitatively fitted by a semi-numerical function~\cite{vadawale_czt_mutau}. 
The same method, with inclusion of an additional effect of charge 
sharing, was used to fit the CZTI spectra~\cite{tanmoy_mutau}. Spectra at three energies
59.54, 88.0 and 122.0 keV from each pixel at every temperature were 
fitted simultaneously to obtain pixel wise parameters such as 
mobility-lifetime product ($\mu\tau$) for the charge carriers, 
effective charge sharing radius ($r_0$), energy resolution etc. 
Figure~\ref{mutau_fit} show the typical fit at three energies. Since 
such simultaneous fitting of the semi-numerical function takes a few 
minutes for one pixel, fitting for the whole data comprising of about 
80000 spectra was carried out using the high performance computing 
facility $Vikram-100$ at Physical Research Laboratory, Ahmedabad. 
Figure~\ref{mutau_param} shows the distribution of the four key
parameters $\mu\tau_e$, $\mu\tau_h$, $r_0$ and $\sigma$ for all pixels. 
These parameters are stored in the CALDB and are used to generate 
the response matrix of the CZTI, full details of which is beyond the 
scope of this paper and will be presented elsewhere.

%%%%%%%%%%%%%%%%% In-orbit operations %%%%%%%%%%%%%%%%%%%%%%%%%%%%%%%%%

\section{In-orbit operations}

\label{inorbit}
CZTI was the first of the five scientific instruments on-board {\astrosat} 
to be switched on. 
After the absolutely flawless launch on 28th September 2015, the first four
days were devoted for monitoring the health and performance of various
spacecraft subsystems and all of them were found to be functioning 
normally. 

%After the launch, first four days were dedicated for monitoring the performance of various sub-systems 
%of the satellite. CZT Imager was the first instrument to be switched on \textcolor{red}{as it is least susceptible to 
%outgassing}. 

On the fifth day after the launch, CZTI processing electronics was powered 
on and the instrument health was monitored for one day. All four quadrants 
were powered-on one-by-one on the following day. All house keeping parameters 
were found to be nominal in-orbit and continues to be so till date. The 
detector temperatures were also found to be within the design limit.

Subsequent three days were devoted for adjusting the detector energy 
thresholds and disabling noisy pixels. Pixels which were found to be 
consistently noisy during the ground calibration and testing were 
already disabled before launch. Some additional pixels were found to
be noisy after switching on the high voltage. These were identified and
quickly disabled. A stable operation of all four quadrants was achieved 
within two days. All detector modules were set to the energy threshold 
of 40 keV prior to launch as a precautionary measure. This was 
gradually brought down to 20 keV during the initial days. Later in 
February 2016 thresholds of large fraction of modules were brought down to 
15 keV. SAA entry and exit operations were initially being carried out by
time tagged ground command, however, later these were automated based on 
the flag from the CPM and since then these operations are being performed 
without any fault. On day nine after the launch, {\astrosat} was pointed to 
its first astrophysical target, the Crab nebula. 

\begin{figure}
\begin{center}
\includegraphics[width=0.5\textwidth]{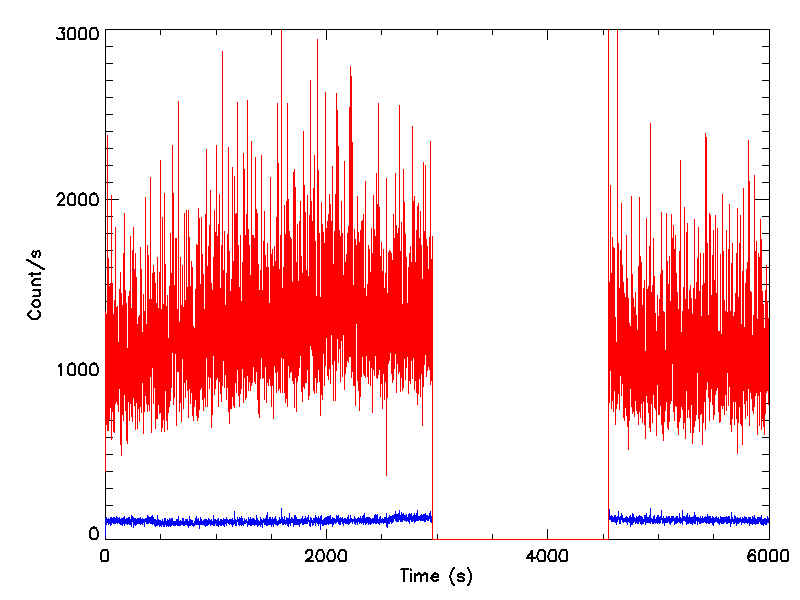}
\caption{ \label{bunch_clean} Raw count rate observed by CZTI (red) and the count rate
after `bunch-cleaning' (blue) demonstrating the effect of the `bunched'
events and effectiveness of the cleaning algorithm.}
\end{center}
\end{figure}

\subsection{CZTI `bunch-cleaning'}

One important observation immediately after switching on the detector
HV, was that the observed count rate was much higher than expected.
It was further observed that the variance of the observed light curve
was much higher than that expected from a Poissonian distribution, 
indicating that these events were inter-related and not truly random. 
On a closer inspection it was found that most of these events occur 
within a very short duration, typically less then 100 micro-second.
It was observed that most of the events are thus `bunched' in time,
but these `bunches' themselves are random in time. With a detailed
analysis of the `bunched' events in terms of both spatial and temporal
correlation, these were postulated to be due to the secondaries from a
high energy charged particle interaction. 
%Hard X-ray detectors are sensitive to charged particle interactions.  
%Particles interacting with the detector lose energy continuously and 
%produce tracks. 
High energy charged particles interacting with the instrument or  
spacecraft body typically produce a shower of secondary particles and 
X-ray photons. As CZTI is composed of pixelated detectors, each pixel 
acts as an independent X-ray detector and thus these secondaries can be
detected as separate events but very closely spaced in time. Thus one 
charged particle can produce events in many pixels of CZTI at the 
same time. 
%We call these multi-hit events as `bunches', as they are 
%bunched  in  time. All these multiple events generated by one particle 
%are being recorded separately and thus huge volume of data was taken by 
%this (but still less than the on-board storage limit).  
Once the origin of the high count rate was understood, we developed 
appropriate algorithms to identify such multi-hit events and clean 
the event data. The data reduction pipeline was suitably modified to 
include these changes. Figure~\ref{bunch_clean} show the effect of 
the `bunch-cleaning' algorithm. It can be seen that after cleaning 
the variance in counts is comparable to Poisson level.

One adverse side effect of the bunched events was that they were 
consuming large data volume which was some times touching the on-board
data storage limit for CZTI. These were also increasing the overall
volume requirements for the later data analysis as well as storage.
Hence after detailed deliberations, it was decided to suppress these
events on-board. This was made possible due to the flexibility of the
CZTI firmware as well as feasibility of on-board reprogramming. A 
software patch was developed to identify and reject the `bunched' events.
After extensive testing of the patch on ground with qualification model 
of CZTI PE, it was uploaded in February 2016 to modify the the on-board 
software on CZTI Processing Electronics. Since then the modified
software has been functioning without any issue and the total data 
volume has been reduced by a factor of four.

%CZTI background is slowly varying within an orbit. Owing to the low 
%inclination of orbit, satellite doesn't pass through deep SAA regions 
%and hence the background variation is less than 50\%. 

\begin{figure}
   \begin{center}
   \begin{tabular}{c c}
   \subfloat[]{\label{alpha_spec} \includegraphics[width=0.45\textwidth]{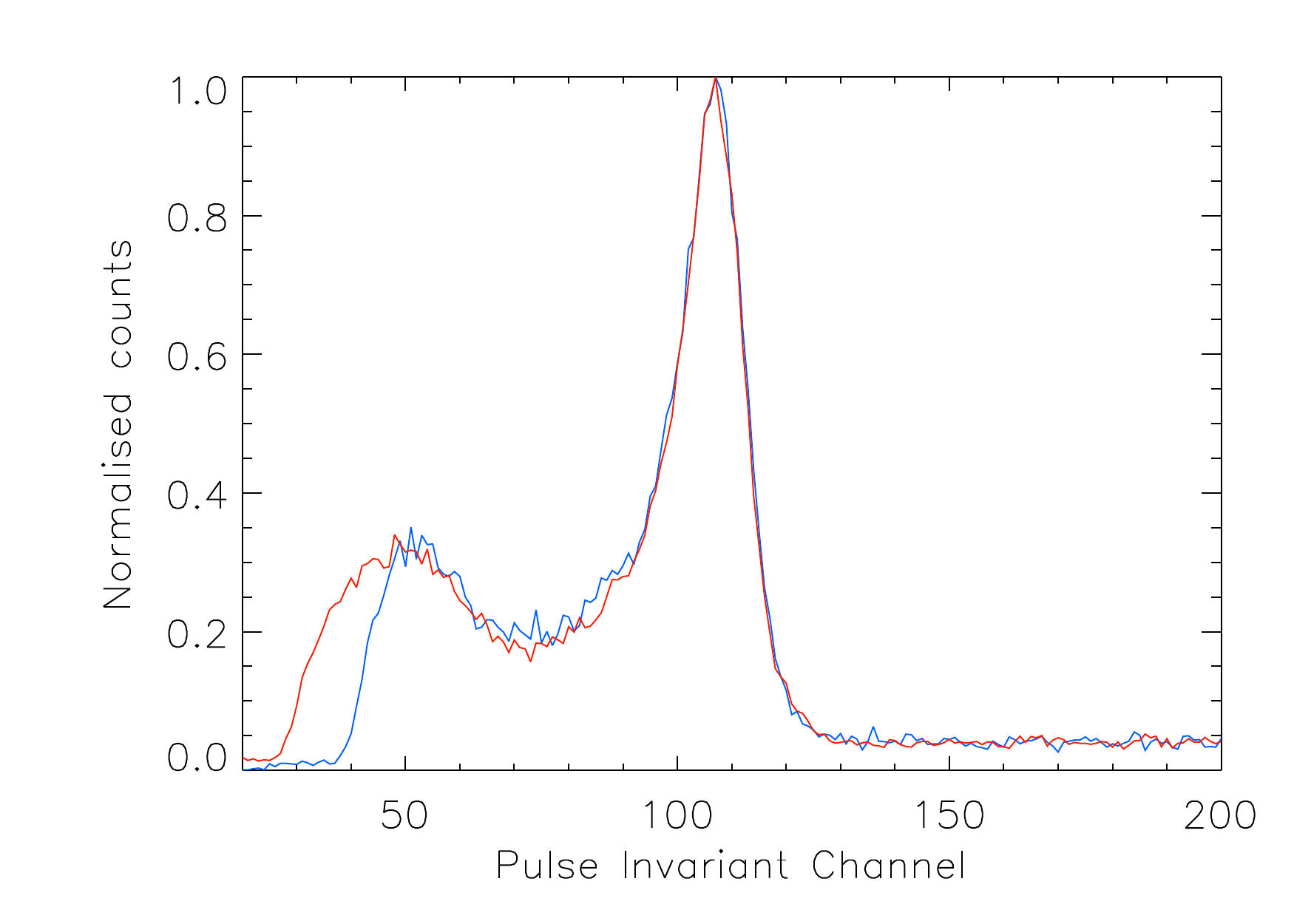}}
   \subfloat[]{\label{alpha_stab} \includegraphics[width=0.45\textwidth]{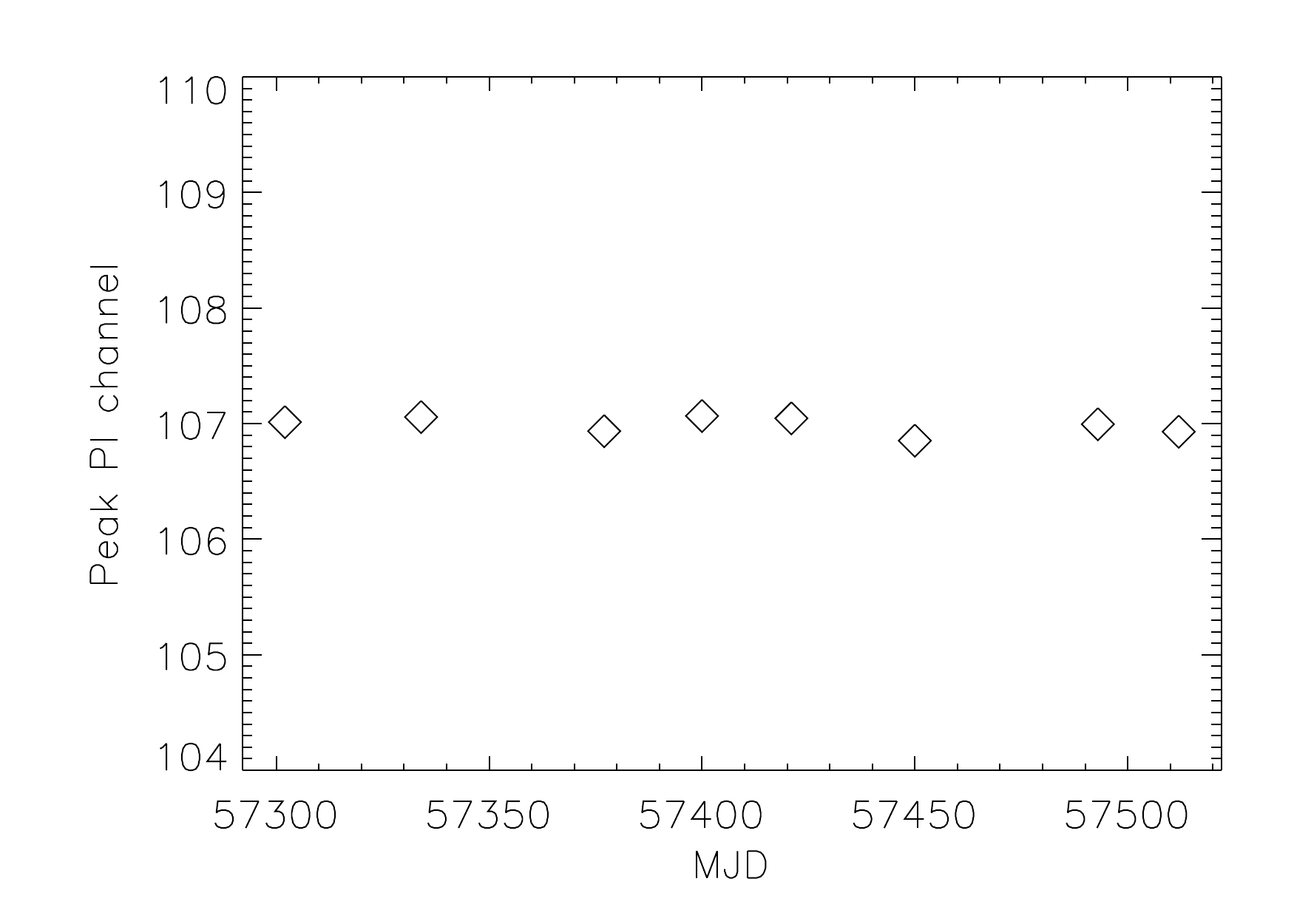}}
   \end{tabular}
    \end{center}
   \caption[example]
   { \label{alphaspec}
   In-flight gain monitoring: (a) Spectrum of 60 keV photons from on-board calibration source for two observations in 
   October 2015 (blue) and May 2016 (red). (b) Peak channel of calibration source spectrum as function of time. %There is no variation in gain. 
}
\end{figure}

\subsection{On-board Gain monitoring}

CZT Imager has an on-board calibration source for continuous monitoring 
of detector gain. The calibration source illuminates the detectors with 
59.54 keV photons, which are identified with simultaneous alpha-tag. Along 
with this fluorescence lines of Tantalum (K$\alpha$ line at 57.5 keV and 
K$\beta$ line at 65.2 keV) provides the in-flight gain calibration. 
In order to check any gain variation over time, the Alpha-tagged events 
from all observations are extracted and examined. Figure~\ref{alpha_spec} 
shows the spectra from on-board calibration source for two different 
observations separated by approximately eight months, 
which look nearly identical except for the difference in detector thresholds. Figure~\ref{alpha_stab} show more comprehensive
analysis of the alpha-tagged events extracted from large number of 
observations where the fitted peak channel is plotted as a function of time.
It can be seen that, so far the gain of CZTI detectors has not changed 
significantly.  

%%%%%%%%%%%%%%%%%%%% Initial results %%%%%%%%%%%%%%%%%%%%%%%%%%%%%%%%%%%%%%%%%%%%%%

\section{Initial results}

\label{results}
The first six month period of {\astrosat} observation was designated for 
Performance Verification (PV) and calibration of all the payloads. 
Observations for performance verification of CZT Imager during these 
six months were designed to characterize various aspects of the instrument 
like effective area, coded-mask response, imaging capability, timing 
capability and off-axis response. Primary target for these observations 
was the Crab nebula which is widely used as standard candle for hard 
X-ray instruments. PV  phase observations also included other bright 
sources Cygnus X-1, GRS~1915+105 and Cygnus X-3. Some initial results 
from these observations are presented here.

\subsection{Crab}

\begin{figure}[h!]
   \begin{center}
   \begin{tabular}{c c c} %% tabular useful for creating an array of images 
   \subfloat[] {\label{crab_image}\includegraphics[width=0.35\textwidth]{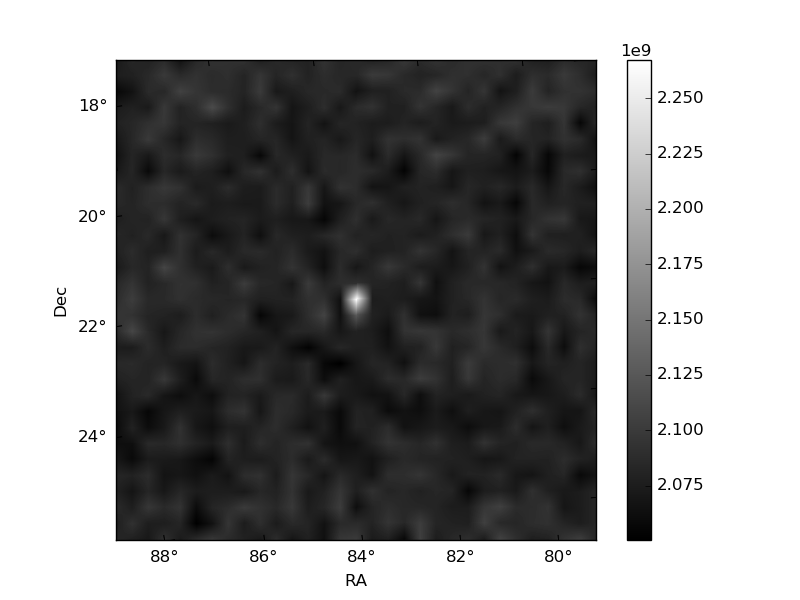}}
   \subfloat[]{ \label{crab_pulse} \includegraphics[width=0.35\textwidth]{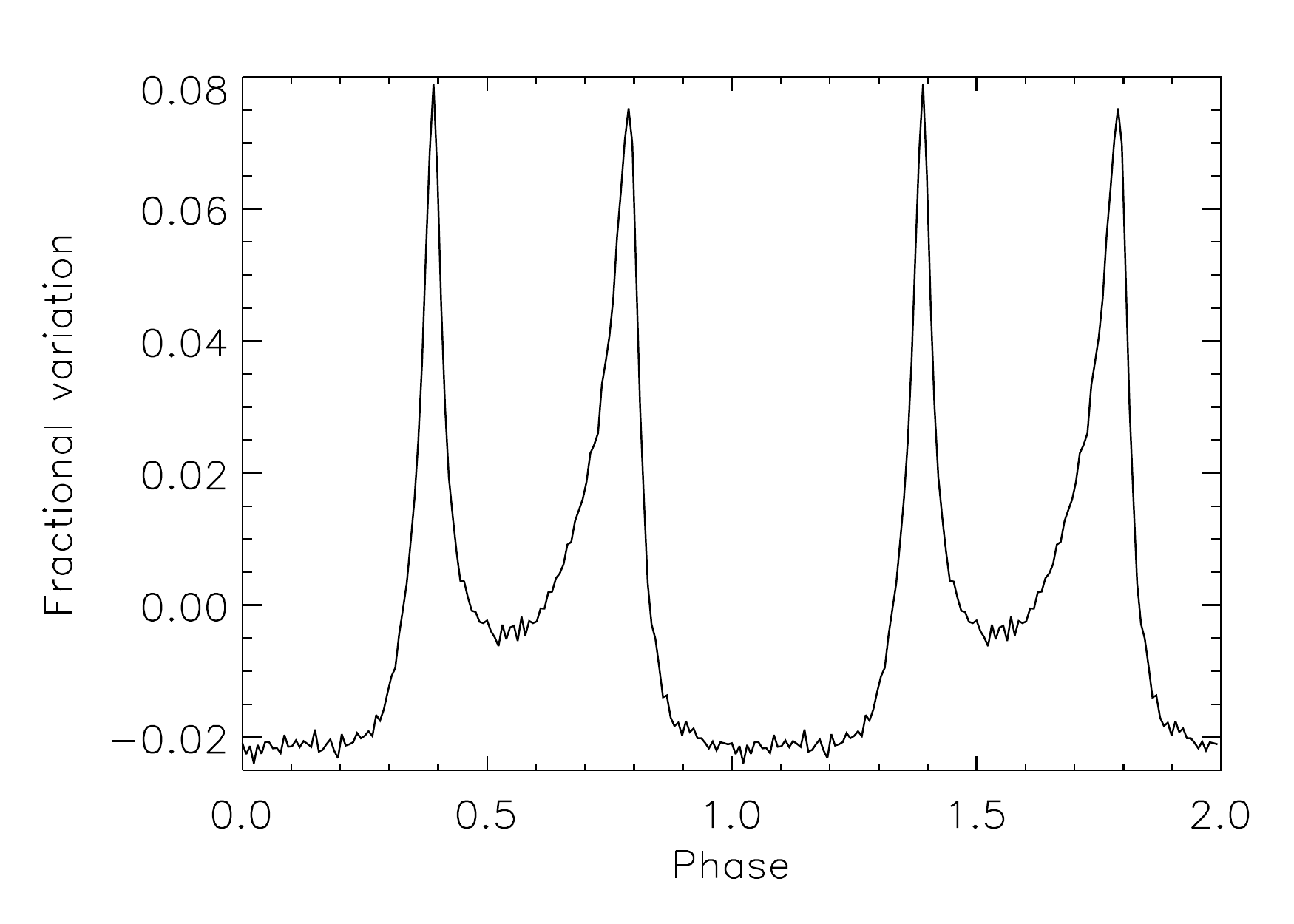}}\\
   \subfloat[]{\label{crab_spec}\includegraphics[width=0.60\textwidth]{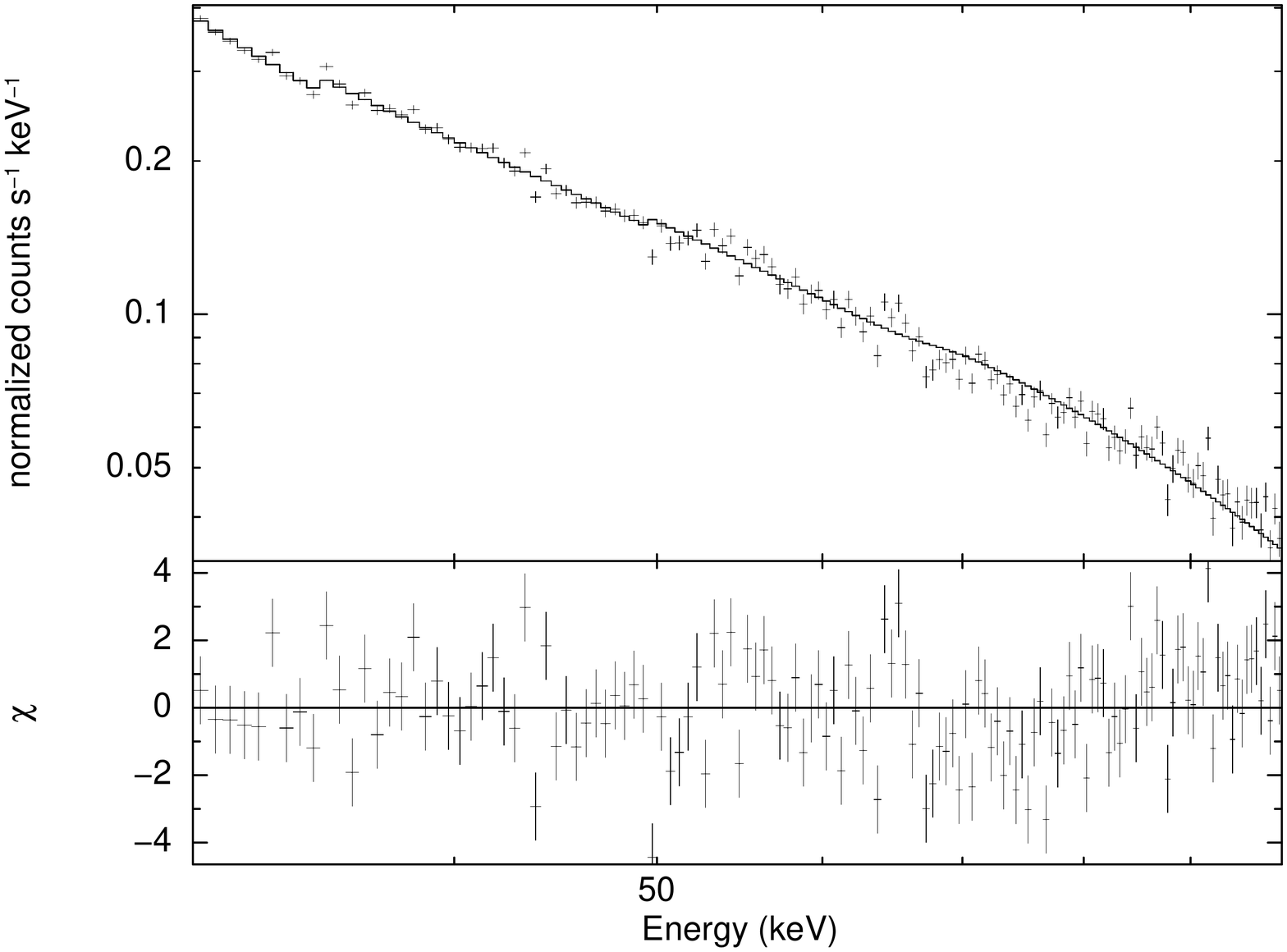}}
   \end{tabular}
   \end{center}
   \caption[example]
   { \label{crab}
    (a) Image of crab obtained from CZTI; (b) Crab pulsar pulse profile; (c) Crab spectrum fitted with a power-law model
}
\end{figure}

Crab being the standard candle in hard X-rays has been observed several 
times by {\astrosat} during the PV phase. Crab was the first source to be 
imaged by CZTI. CZTI employs indirect imaging where image is obtained by 
deconvolution of detector plane histogram with the mask pattern. There 
are several image reconstruction methods and imaging by FFT is much faster 
compared to the other methods. Figure ~\ref{crab_image} shows the image of Crab 
from CZTI, reconstructed using FFT method. 
%\textcolor{red}{Imaging resolution, what number to quote???}.   
Observations were made with crab at different positions in the field of 
view to characterize the coded-mask response and imaging performance. 
These are analyzed to obtain the accurate alignment between the mask 
and detector pixels. 

CZTI records time stamps of photons with an accuracy of 20$~{\mu}s$, and 
thus allows high time resolution studies. The Crab nebula also hosts the 
Crab pulsar which has pulse period of 33.5 milli-seconds and thus is a 
good candidate for establishing the timing capabilities of CZTI. Crab 
pulse profile generated from barycentric corrected event time stamps is 
shown in figure~\ref{crab_pulse}. The pulse profile matches well with 
observations from other instruments in the same energy range. Preliminary 
analysis with radio ephemeris suggests that absolute time accuracy of CZTI 
is within 200$~{\mu}s$. Simultaneous observations with GMRT and INTEGRAL 
is available for some of the PV observations of crab. Detailed analysis 
is under progress.

Spectral response for CZTI is generated based on modeling extensive ground 
calibration data. A model based on charge trapping and charge sharing is 
developed for generation of redistribution matrix. Crab spectrum is used 
to verify the response of X-ray instruments. In CZTI, background subtracted 
spectrum is obtained by mask-weighting where photons in each pixel is 
assigned weights based on the mask open fraction. Mask-weighted spectrum 
of Crab for an observation with exposure time of 110 ks is shown in 
figure~\ref{crab_spec}. The spectrum is fitted with canonical crab spectral 
model, power law, with 2\% systematic error. Fitted power law index 2.09 is consistent with widely 
accepted value~\cite{crab_std}, which validates the response matrix of the instrument. 
Effective area estimation using multiple crab observations is under progress. 

\subsection{Gamma Ray Bursts}

At energies beyond 100 keV CZT Imager acts as an open detector. This 
provides CZTI with the capability to monitor transient events like 
Gamma Ray Bursts (GRB). GRBs are most energetic explosions in the 
universe which makes them detectable at cosmological distances. A typical 
GRB spectrum peaks at 100 keV to 1 MeV energy range. Observations with 
CZTI can complement the data from dedicated GRB missions like $Swift$~\cite{swift} 
and $Fermi$~\cite{fermi_gbm,fermi_lat} to provide broad band spectral coverage.

\begin{figure}[h!]
   \begin{center}
   \begin{tabular}{c c} %% tabular useful for creating an array of images 
   \subfloat[] {\label{grb_lc} \includegraphics[width=0.45\textwidth]{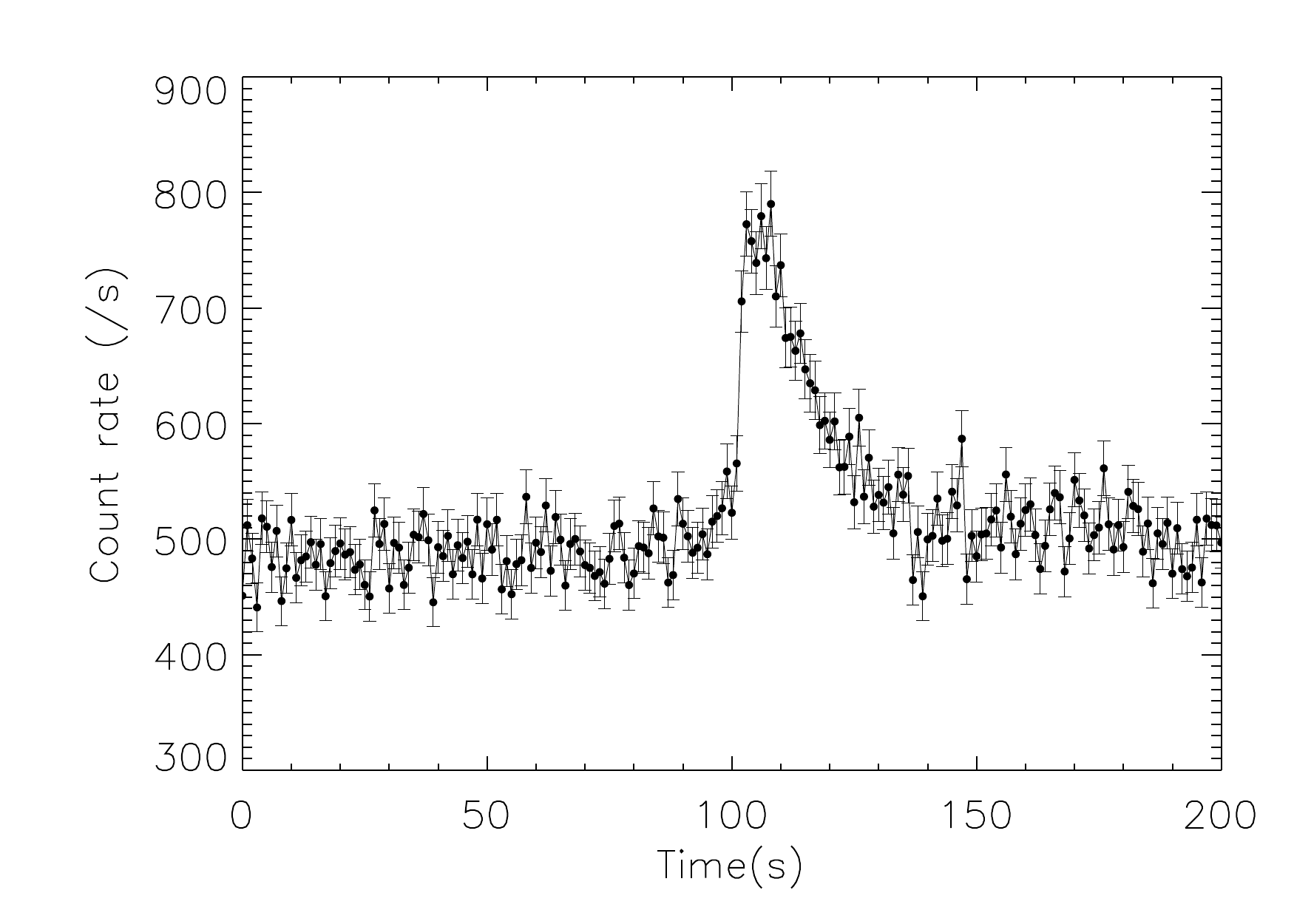}}
   \subfloat[]{\label{grb_image} \includegraphics[width=0.60\textwidth]{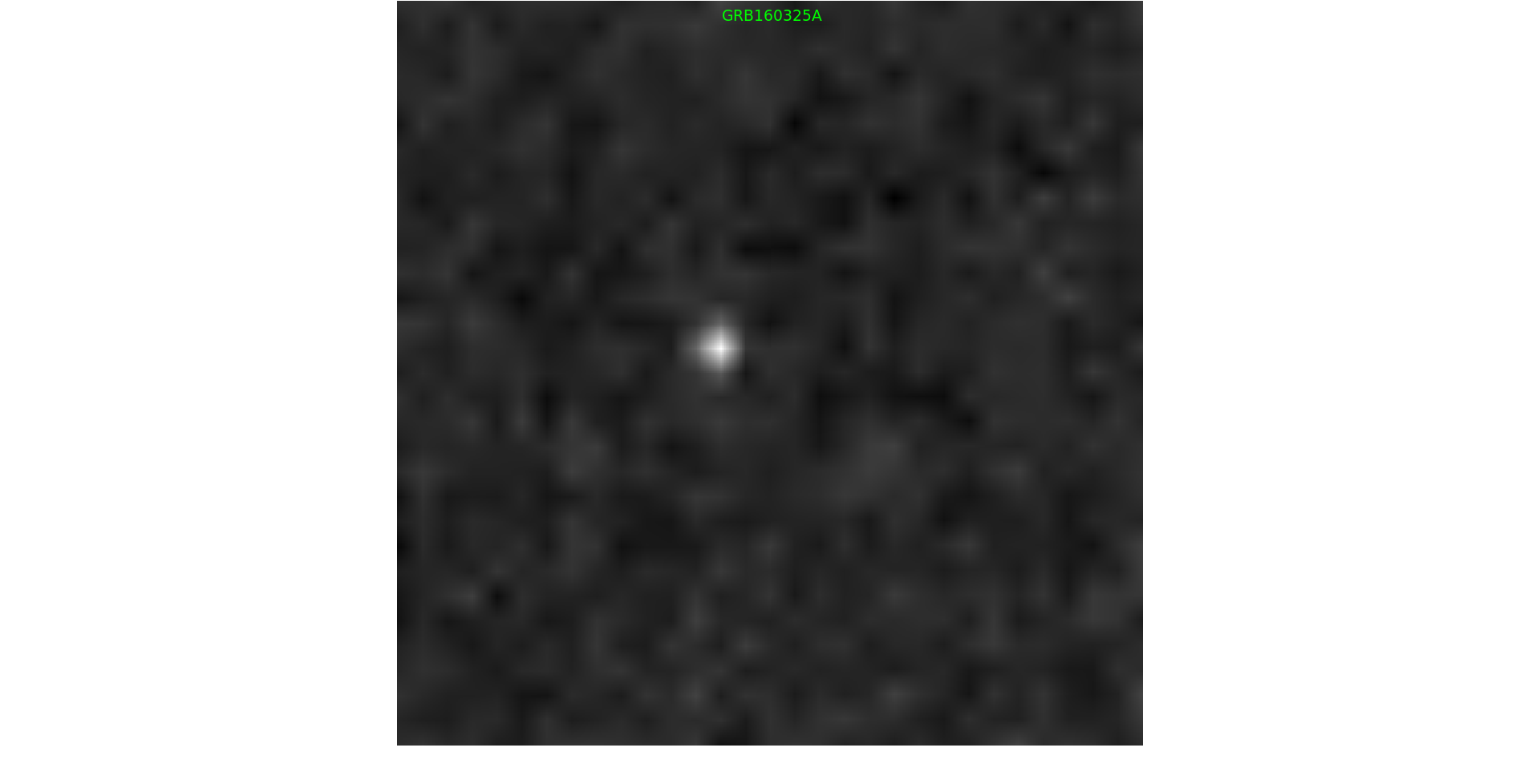}}
   \end{tabular}
   \end{center}
   \caption[example]
   { \label{grbs}
    GRBs with CZTI: (a) Light curve of GRB151006A, the first GRB detected by CZTI; (b) Image of GRB160325A 
detected in primary field of view of CZTI.
}
\end{figure}

Incidentally, on the first day of observations with CZTI, Swift BAT 
reported a GRB at 60.7 degrees off-axis to {\astrosat} pointing direction. 
This GRB (GRB151006A) was clearly detected by CZTI~\cite{GCN1} and the 
light curve is shown in figure~\ref{grb_lc}. Detailed broad-band spectral 
analysis of this GRB is performed and will be reported elsewhere.
Since then CZTI has detected 9 GRBs till May 2016. This includes triggered 
searches and independent detections. One of these GRBs (GRB 160325A) was 
detected within the primary field of view of CZTI. This allowed accurate 
localization of the GRB. Image reconstructed by deconvolution is shown in 
figure~\ref{grb_image}, the source coordinates obtained from CZTI matches 
with the value from Swift. Coarse localization is possible for GRBs which 
occur outside the primary field of view as well. The collimators cast 
shadow on the detector plane which can be used to reconstruct the source 
location.  

\subsection{Cygnus X-1}

It is well known that the wide band X-ray spectra of galactic black hole 
binaries like Cygnus X-1 consist of various spectral components such 
as accretion disk emission, reflection component, Comptonized emission, 
non-thermal power-law etc and the relative contribution of these 
components vary with the spectral state~\cite{rem_mcc}. Resolving 
various components of the emission from such systems require sensitive 
simultaneous broad band spectral measurements at energies ranging from 
$<$1 keV to $>$100 keV. {\astrosat} with its wide band coverage provides 
the right kind of observations required.

\begin{figure}[h!]
   \begin{center}
   \begin{tabular}{c c} %% tabular useful for creating an array of images 
   \subfloat[]{\label{cygx1_img} \includegraphics[width=0.42\textwidth]{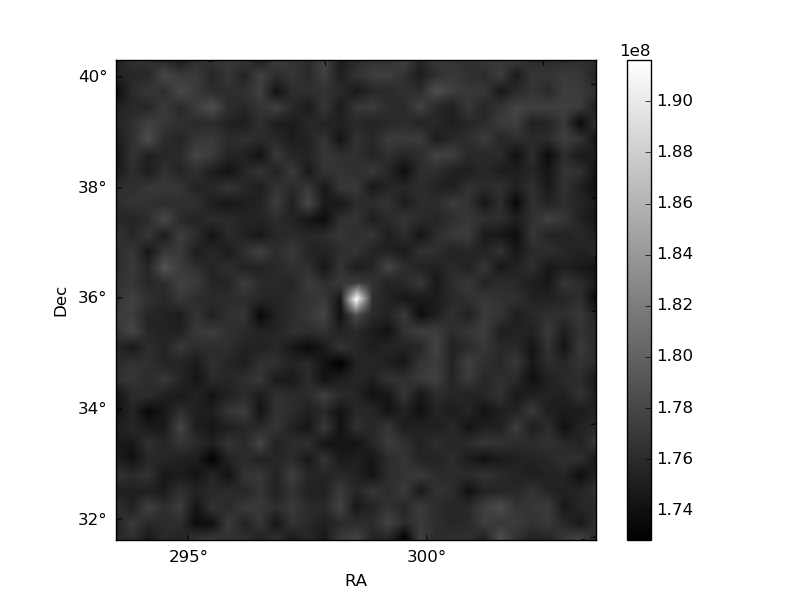}}
   \subfloat[]{\label{cygx1_bat} \includegraphics[width=0.40\textwidth]{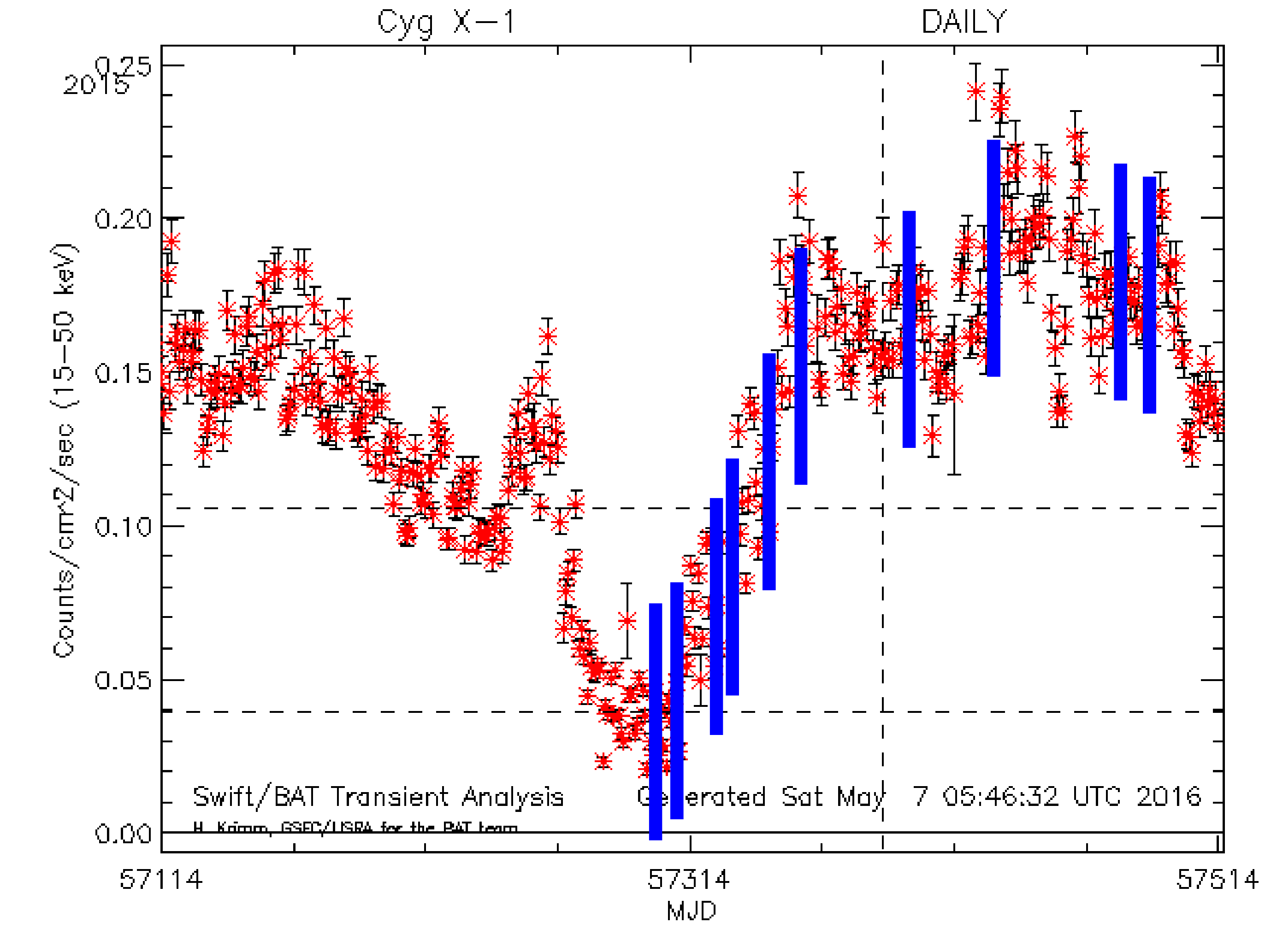}}
   \end{tabular}
   \end{center}
   \caption[example]
   { \label{cyg_x1}
    Cygnus X-1: (a) CZTI Image of Cygnus X-1 in hard state; (b) Swift BAT light curve of Cygnus X-1 for 400 days.
Vertical lines correspond to {\astrosat} observations. {\astrosat} has well sampled observations during
transition from soft state to hard state.
}
\end{figure}

The galactic micro quasar Cygnus X-1 has been observed several times 
during the PV phase of {\astrosat}. In figure~\ref{cygx1_img}, CZTI image 
of Cygnus X-1 in its hard state is shown. Figure~\ref{cygx1_bat} shows 
the Swift BAT monitoring light curve of Cygnus X-1 for about 400 days 
and {\astrosat} observation is marked by the vertical lines. Incidentally 
{\astrosat} CZTI started its observations when the source was in very peculiar 
ultra soft state with hard X-ray flux of $\sim$150 mCrab. In the last 
eight months, Cygnus X-1 has undergone transition from ultra soft state 
to hard state and {\astrosat} has nicely covered the whole state transition 
by frequent pointed observations. Spectra from CZTI along with other 
{\astrosat} instruments will provide valuable information on the changes in
the radiative processes and the accretion geometry during such state 
transitions. For some of these observations of {\astrosat}, 
simultaneous observation by NuStar is also available. Also contemporaneous  
observations at other wavebands like Infra-red (from Mt. Abu Infrared 
Observatory) and and Radio (from GMRT) are also available. 
Detailed analysis is under progress.

\begin{figure}[h!]
   \begin{center}
   %\begin{tabular}{c c} %% tabular useful for creating an array of images 
   %\subfloat[]{
   \includegraphics[width=0.60\textwidth]{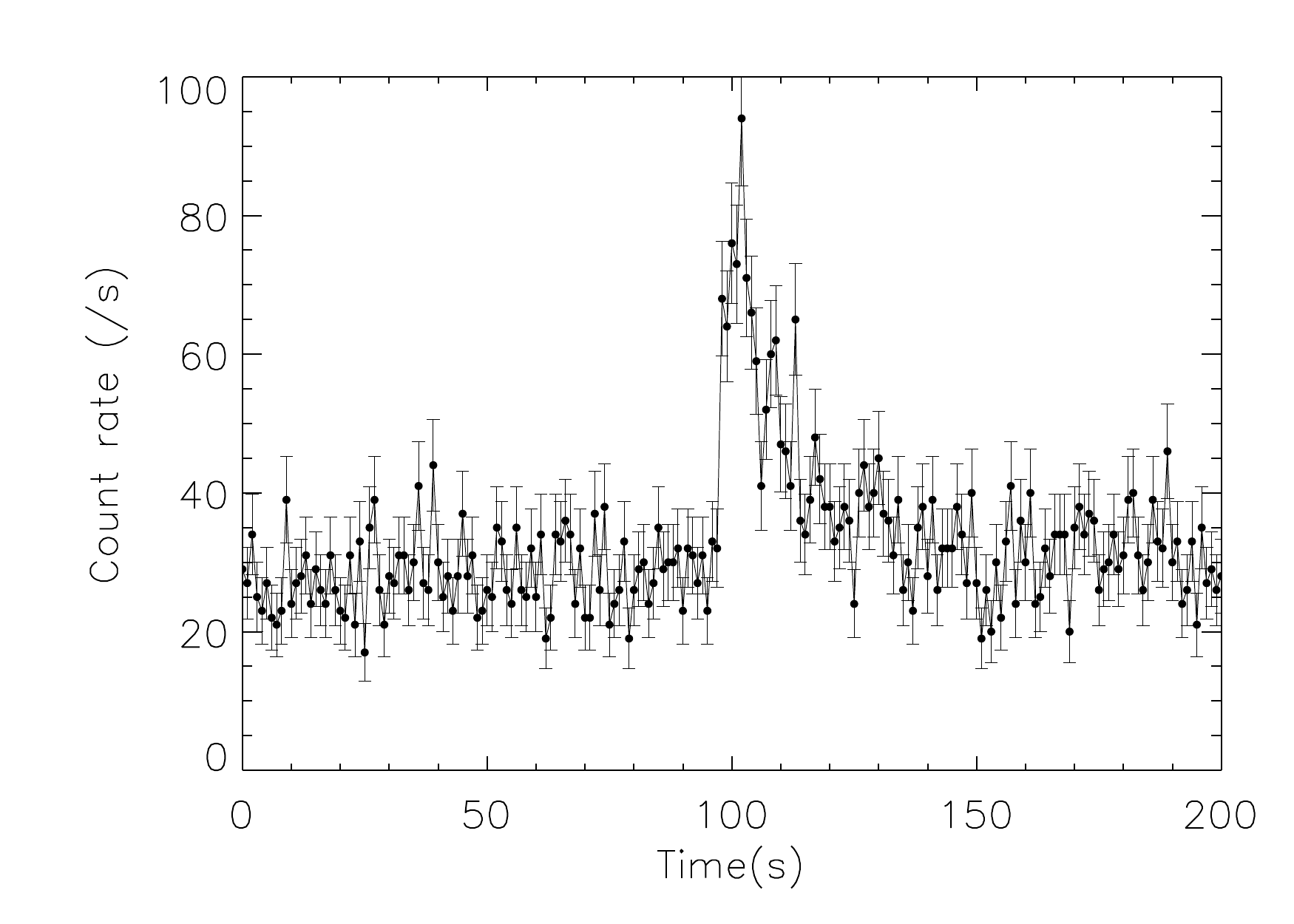}
   %\end{tabular}
   \end{center}
   \caption[example]
   { \label{dblevt}
	Light curve of Compton scattered double events for GRB160131A. 
}
\end{figure}

\subsection{Polarimetry}

X-ray polarimetry is a largely unexplored area in the field of X-ray 
astronomy. Thus exploring any avenue which can provide polarimetric 
measurements is of great importance. CZTI having pixelated detectors 
is sensitive to the Compton scattered events which allows polarization 
measurements in the energy range of 100-300 keV. Polarimetric capability 
of CZTI has been explored in detail using Monte-Carlo simulations
~\cite{tanmoy_cztpol} and has also been experimentally verified on 
ground with the flight module. Detailed study of polarimetric sensitivity  
shows that sensitive polarization measurements of bright sources like 
Crab and Cygnus X-1 is possible with CZTI~\cite{vadawale_cztpol}.

Gamma Ray Bursts being very bright, are found very useful to validate 
the criteria applied in the selection of Compton scattered events in CZTI. 
Figure~\ref{dblevt} shows the light curve of Compton scattered double 
events for GRB160131A. This validates the criterion used in the selection 
of double events and is a firm demonstration of detection of Compton 
events from an astrophysical source. Azimuthal distribution of selected 
Compton events show clear modulation~\cite{GCN2}. Further work is under 
progress to perform detailed mass-modeling of the instrument and satellite 
which is required to quantify the degree and angle of polarization. 
Crab and Cygnus X-1 are the other candidates for hard X-ray polarization 
measurement with CZTI. Both these sources have been extensively observed
by {\astrosat}. Results from the preliminary analysis of data for these 
sources are very promising and will be reported elsewhere after more 
detailed scrutiny.

%%%%%% SUMMARY %%%%%%%%%%%%%%%%%%%

\section{Summary}

    The Cadmium-Zinc-Telluride Imager (CZTI) on-board {\astrosat} is designed for 
simultaneous hard X-ray imaging and spectroscopy in energy range of 20 - 150 keV. 
The full instrument, in terms of both hardware and software, is functioning exactly 
as designed and is performing very satisfactorily. The enhanced capabilities of 
CZTI i.e. hard X-ray monitoring and polarimetry in the extended energy range of 
100 - 300 keV, are also functioning as per expectation. Detailed characterization 
of the CZTI imaging, spectroscopic and timing performance has been carried out 
during the first six months of performance verification observations. 
%{\bf The spectral response of CZTI has been found to be stable.}
During the initial observations, it was found that CZTI is highly sensitive to the 
`particle shower' like events due to the large number of pixels, 
resulting in much larger data volume mostly consisting of events 
useless for the scientific objectives of CZTI. A remedy for this problem 
has been implemented in the onboard software as well as in the CZTI data 
processing pipeline. The full data analysis software, including the 
calibration database (CALDB) and the higher level data products 
generation software, is fully functional and is released for wider 
use/testing. Overall the CZTI instrument is 
performing as per the expectations.
%CZTI on-board {\astrosat} is working flawless. Six months of performance verification 
%phase is completed in March 2016 and currently Guaranteed time science observations are going on.
%The data analysis pipeline is now completely functional. First version of calibration database (CALDB) 
%is also generated. Spectral and imaging calibration is under progress and would be completed soon.
 
\acknowledgments

The CZT-Imager is designed and developed with the help of a large 
number of organisations and teams; the basic design conceptualization, 
initial development and overall project coordination was carried out by
TIFR group, the coded mask design and imaging algorithms were developed 
by the IUCAA group, instrument mechanical and thermal design was carried 
out by ISRO Satellite Center (ISAC), assembly and testing was carried 
out at the Avionics group of Vikram Sarabhai Space Center (VSSC), the 
pipe-line software was developed at Space Application Center (SAC), the 
CALDB and the polarimetry algorithms were developed at Physical Research 
Laboratory (PRL). The CZTI payload operation center is hosted by IUCAA. 
The authors would particularly like to acknowledge the help received 
from the {\astrosat} mission team at ISRO in all aspects of the CZTI 
realisation and operation. We would also like to acknowledge the support
from the Computer center at PRL while using the high performance 
computing facility Vikram 100. 

\bibliography{reference_list} % bibliography data in report.bib

\begin{thebibliography}{10}

\bibitem{singh_astrosat}
{Singh}, K.~P., {Tandon}, S.~N., {Agrawal}, P.~C., {Antia}, H.~M., {Manchanda},
  R.~K., {Yadav}, J.~S., {Seetha}, S., {Ramadevi}, M.~C., {Rao}, A.~R.,
  {Bhattacharya}, D., {Paul}, B., {Sreekumar}, P., {Bhattacharyya}, S.,
  {Stewart}, G.~C., {Hutchings}, J., {Annapurni}, S.~A., {Ghosh}, S.~K.,
  {Murthy}, J., {Pati}, A., {Rao}, N.~K., {Stalin}, C.~S., {Girish}, V.,
  {Sankarasubramanian}, K., {Vadawale}, S., {Bhalerao}, V.~B., {Dewangan},
  G.~C., {Dedhia}, D.~K., {Hingar}, M.~K., {Katoch}, T.~B., {Kothare}, A.~T.,
  {Mirza}, I., {Mukerjee}, K., {Shah}, H., {Shah}, P., {Mohan}, R., {Sangal},
  A.~K., {Nagabhusana}, S., {Sriram}, S., {Malkar}, J.~P., {Sreekumar}, S.,
  {Abbey}, A.~F., {Hansford}, G.~M., {Beardmore}, A.~P., {Sharma}, M.~R.,
  {Murthy}, S., {Kulkarni}, R., {Meena}, G., {Babu}, V.~C., and {Postma}, J.,
  ``{ASTROSAT mission},'' in [{\em Space Telescopes and Instrumentation 2014:
  Ultraviolet to Gamma Ray}{\nolinebreak\hspace{0.1em}]},  {\em \procspie} {\bf
  9144},  91441S (July 2014).

\bibitem{rao_alpha}
{Rao}, A.~R., {Naik}, S., {Patil}, M., {Malkar}, J.~P., and {Kalyan Kumar},
  R.~P.~S., ``{An alpha tagged X-ray source for the calibration of space borne
  X-ray detectors},'' {\em Nuclear Instruments and Methods in Physics Research
  A}~{\bf 616},  55--58 (Apr. 2010).

\bibitem{vadawale_czt_mutau}
{Vadawale}, S.~V., {Shanmugam}, M., {Purohit}, S., {Acharya}, Y.~B., and
  {Sudhakar}, M., ``{Experimental measurements of charge carrier mobility:
  lifetime products for large sample of pixilated CZT detectors},'' in [{\em
  High Energy, Optical, and Infrared Detectors for Astronomy
  V}{\nolinebreak\hspace{0.1em}]},  {\em \procspie} {\bf 8453},  84532K (July
  2012).

\bibitem{tanmoy_mutau}
{Chattopadhyay}, T., {Vadawale}, S.~V., {Rao}, A.~R., {Bhattacharya}, D.,
  {Mithun}, N.~P.~S., and {Bhalerao}, V.~B., ``{Line profile modelling for
  multi-pixel CZT detectors},'' in [{\em In this
  Volume}{\nolinebreak\hspace{0.1em}]},  {\em \procspie} (2016).

\bibitem{crab_std}
{Kirsch}, M.~G., {Briel}, U.~G., {Burrows}, D., {Campana}, S., {Cusumano}, G.,
  {Ebisawa}, K., {Freyberg}, M.~J., {Guainazzi}, M., {Haberl}, F., {Jahoda},
  K., {Kaastra}, J., {Kretschmar}, P., {Larsson}, S., {Lubi{\'n}ski}, P.,
  {Mori}, K., {Plucinsky}, P., {Pollock}, A.~M., {Rothschild}, R., {Sembay},
  S., {Wilms}, J., and {Yamamoto}, M., ``{Crab: the standard x-ray candle with
  all (modern) x-ray satellites},'' in [{\em UV, X-Ray, and Gamma-Ray Space
  Instrumentation for Astronomy XIV}{\nolinebreak\hspace{0.1em}]},  {Siegmund},
  O.~H.~W., ed., {\em \procspie} {\bf 5898},  22--33 (Aug. 2005).

\bibitem{swift}
{Gehrels}, N., {Chincarini}, G., {Giommi}, P., {Mason}, K.~O., {Nousek}, J.~A.,
  {Wells}, A.~A., {White}, N.~E., {Barthelmy}, S.~D., {Burrows}, D.~N.,
  {Cominsky}, L.~R., {Hurley}, K.~C., {Marshall}, F.~E., {M{\'e}sz{\'a}ros},
  P., {Roming}, P.~W.~A., {Angelini}, L., {Barbier}, L.~M., {Belloni}, T.,
  {Campana}, S., {Caraveo}, P.~A., {Chester}, M.~M., {Citterio}, O., {Cline},
  T.~L., {Cropper}, M.~S., {Cummings}, J.~R., {Dean}, A.~J., {Feigelson},
  E.~D., {Fenimore}, E.~E., {Frail}, D.~A., {Fruchter}, A.~S., {Garmire},
  G.~P., {Gendreau}, K., {Ghisellini}, G., {Greiner}, J., {Hill}, J.~E.,
  {Hunsberger}, S.~D., {Krimm}, H.~A., {Kulkarni}, S.~R., {Kumar}, P.,
  {Lebrun}, F., {Lloyd-Ronning}, N.~M., {Markwardt}, C.~B., {Mattson}, B.~J.,
  {Mushotzky}, R.~F., {Norris}, J.~P., {Osborne}, J., {Paczynski}, B.,
  {Palmer}, D.~M., {Park}, H.-S., {Parsons}, A.~M., {Paul}, J., {Rees}, M.~J.,
  {Reynolds}, C.~S., {Rhoads}, J.~E., {Sasseen}, T.~P., {Schaefer}, B.~E.,
  {Short}, A.~T., {Smale}, A.~P., {Smith}, I.~A., {Stella}, L., {Tagliaferri},
  G., {Takahashi}, T., {Tashiro}, M., {Townsley}, L.~K., {Tueller}, J.,
  {Turner}, M.~J.~L., {Vietri}, M., {Voges}, W., {Ward}, M.~J., {Willingale},
  R., {Zerbi}, F.~M., and {Zhang}, W.~W., ``{The Swift Gamma-Ray Burst
  Mission},'' {\em \apj}~{\bf 611},  1005--1020 (Aug. 2004).

\bibitem{fermi_gbm}
{Meegan}, C., {Lichti}, G., {Bhat}, P.~N., {Bissaldi}, E., {Briggs}, M.~S.,
  {Connaughton}, V., {Diehl}, R., {Fishman}, G., {Greiner}, J., {Hoover},
  A.~S., {van der Horst}, A.~J., {von Kienlin}, A., {Kippen}, R.~M.,
  {Kouveliotou}, C., {McBreen}, S., {Paciesas}, W.~S., {Preece}, R., {Steinle},
  H., {Wallace}, M.~S., {Wilson}, R.~B., and {Wilson-Hodge}, C., ``{The Fermi
  Gamma-ray Burst Monitor},'' {\em \apj}~{\bf 702},  791--804 (Sept. 2009).

\bibitem{fermi_lat}
{Atwood}, W.~B., {Abdo}, A.~A., {Ackermann}, M., {Althouse}, W., {Anderson},
  B., {Axelsson}, M., {Baldini}, L., {Ballet}, J., {Band}, D.~L.,
  {Barbiellini}, G., and et~al., ``{The Large Area Telescope on the Fermi
  Gamma-Ray Space Telescope Mission},'' {\em \apj}~{\bf 697},  1071--1102 (June
  2009).

\bibitem{GCN1}
{Bhalerao}, V., {Bhattacharya}, D., {Rao}, A.~R., and {Vadawale}, S., ``{GRB
  151006A: Astrosat CZTI detection.},'' {\em GCN Circ.}~{\bf 18422} (2015).

\bibitem{rem_mcc}
{Remillard}, R.~A. and {McClintock}, J.~E., ``{X-Ray Properties of Black-Hole
  Binaries},'' {\em \araa}~{\bf 44},  49--92 (Sept. 2006).

\bibitem{tanmoy_cztpol}
{Chattopadhyay}, T., {Vadawale}, S.~V., {Rao}, A.~R., {Sreekumar}, S., and
  {Bhattacharya}, D., ``{Prospects of hard X-ray polarimetry with
  Astrosat-CZTI},'' {\em Experimental Astronomy}~{\bf 37},  555--577 (Nov.
  2014).

\bibitem{vadawale_cztpol}
{Vadawale}, S.~V., {Chattopadhyay}, T., {Rao}, A.~R., {Bhattacharya}, D.,
  {Bhalerao}, V.~B., {Vagshette}, N., {Pawar}, P., and {Sreekumar}, S., ``{Hard
  X-ray polarimetry with Astrosat-CZTI},'' {\em \aap}~{\bf 578},  A73 (June
  2015).

\bibitem{GCN2}
{Vadawale}, S.~V., {Chattopadhyay}, T., {Mithun}, N.~P.~S., {Rao}, A.~R.,
  {Bhattacharya}, D., and {Bhalerao}, V., ``{GRB160131A: detection of
  polarisation by Astrosat CZTI.},'' {\em GCN Circ.}~{\bf 19011} (2016).

\end{thebibliography}
\bibliographystyle{spiebib} % makes bibtex use spiebib.bst
 
\end{document}